\documentclass[journal=nalefd,manuscript=letter]{achemso}

\usepackage[version=3]{mhchem} 
\usepackage{graphicx}

\author{Daniel Wegner}
\email{daniel.wegner@uni-muenster.de} \affiliation[Universit\"at
M\"unster and CeNTech]{Physikalisches Institut and Center for
Nanotechnology (CeNTech), Westf\"alische Wilhelms-Universit\"at
M\"unster, 48149 M\"unster, Germany} \affiliation[University of
California at Berkeley]{Department of Physics, University of
California at Berkeley, and Materials Science Division, Lawrence
Berkeley National Laboratory, Berkeley, California 94720-7300, USA}

\author{Ryan Yamachika}
\affiliation[University of California at Berkeley]{Department of
Physics, University of California at Berkeley, and Materials Science
Division, Lawrence Berkeley National Laboratory, Berkeley,
California 94720-7300, USA}

\author{Xiaowei Zhang}
\affiliation[University of California at Berkeley]{Department of
Physics, University of California at Berkeley, and Materials Science
Division, Lawrence Berkeley National Laboratory, Berkeley,
California 94720-7300, USA}

\author{Yayu Wang}
\affiliation[University of California at Berkeley]{Department of
Physics, University of California at Berkeley, and Materials Science
Division, Lawrence Berkeley National Laboratory, Berkeley,
California 94720-7300, USA}

\author{Michael F.\ Crommie}
\affiliation[University of California at Berkeley]{Department of
Physics, University of California at Berkeley, and Materials Science
Division, Lawrence Berkeley National Laboratory, Berkeley,
California 94720-7300, USA}

\author{Nicol\'as Lorente}
\affiliation[Centre d'Investigaci\'{o}  en Nanoci\`{e}ncia i
Nanotecnologia]{Centre d'Investigaci\'{o}  en Nanoci\`{e}ncia i
Nanotecnologia, CIN2 (CSIC-ICN), Campus de la UAB, 08193 Bellaterra,
Spain}

\title[Molecular site determination via IETS]{Adsorption site determination of a molecular monolayer via inelastic tunneling}

\keywords{scanning tunneling microscopy, density functional theory,
hybrid organometallic interfaces, adsorption site determination,
inelastic electron tunneling spectroscopy, vibrational modes}

\begin{document}

\begin{abstract}
We have combined scanning tunneling microscopy with inelastic
electron tunneling spectroscopy (IETS) and density functional theory
(DFT) to study a tetracyanoethylene monolayer on Ag(100). Images
show that the molecules arrange in locally ordered patterns with
three non-equivalent, but undeterminable, adsorption sites. While
scanning tunneling spectroscopy only shows subtle variations of the
local electronic structure at the three different positions, we find
that vibrational modes are very sensitive to the local atomic
environment. IETS detects sizeable mode frequency shifts of the
molecules located at the three topographically detected sites, which
permits us to determine the molecular adsorption sites through
identification with DFT calculations.
\end{abstract}

%
%

Many phenomena are determined by the way molecules adsorb on
substrates, e.g., heterogeneous catalysis
\cite{Hofmann1994,Frank2000}, transport in molecular electronics
\cite{Ratner2011}, and competing many-body interactions
\cite{Franke2011}. This has led to a large effort to apply surface
science tools for adsorption analysis. Particularly powerful
techniques for investigating large arrays of ordered molecules use
diffraction and interference effects. For instance, low-energy
electron diffraction \cite{leed} conveys information on molecular
arrangements, X-ray standing waves reveal the adsorbate's vertical
geometrical structure \cite{Hauschild2005}, and surface X-ray
diffraction can determine the structure of molecular adlayers by
counteracting the penetration depth of X-rays using
grazing-incidence angles \cite{sxr}. As the number of molecules
inside the repeating pattern increases, these techniques grow more
complex and the diffraction patterns become increasingly difficult
to unravel. In such cases, local-probe techniques can be
particularly useful. Indeed, the scanning tunneling microscope (STM)
has proven to be a powerful tool to for structure determination of
atomic surfaces and molecular layers
\cite{Binnig1983,deFeyter2003,Heinz2004}. Recent progress in
scanning probe techniques even permits the detection of chemical
bonds within and between molecules \cite{AFMbonds,HydrogenSTM}.

With the active search of molecular nanostructures, very complex
arrays of dense molecular layers have become ubiquitous in surface
science
\cite{Barth2005,CrommieC60Pinwheel,Besenbacher,CuPcGraphene,InvMelt,ThiolateGaAs}.
Structure analysis by diffraction techniques becomes hopeless when
commensurability issues between molecule and substrate set in,
impeding long-range order on the surface. While such structures can
only be studied by local-probe techniques, exact determination of
adsorption sites via STM alone remains very difficult due to the
challenge to acquire molecular and surface atom positions
simultaneously with high precision
\cite{EiglerBenzene1993,Meyer96,Berndt98,Lagoute04,Tautz07}.

Analysis of the scanning tunneling spectroscopy (STS) structure by
inelastic effects during electron tunneling allows a deeper
understanding of molecular behavior at a surface. Such inelastic
electron tunneling spectroscopy (IETS) has become a very successful
tool to detect vibrational, photon, and even spin excitations
\cite{Sti98,Cavar05,Heinrich04}. In its vibrational version, the
extraordinary spatial resolution of IETS \cite{Ho2002} allows
chemical analysis capabilities at the atomic scale
\cite{Lee1999,Pascual2001,Lauhon2000,Kim2002,Weg09,Paulsson2010}.
Moreover, IETS is generally below the meV energy resolution. Hence,
we can explore small environment-induced frequency shifts with
intra-molecular resolution. This naturally leads us to explore
whether IETS can improve adsorption site determination through
observation of systematic shifts in vibrational-mode frequencies for
molecules locally bound at different adsorption sites on a surface.

Here, we demonstrate that IETS can indeed yield accurate information
on molecular binding geometries within a defect-free densely-packed
molecular monolayer that lacks long-range order and thus would be
challenging to study using diffraction-based techniques. The studied
system is tetracyanoethylene (TCNE, \ce{C6N4}) adsorbed on Ag(100).
TCNE is known to be an electron acceptor showing strong interactions
with metals due to the four cyano-group low-energy empty $\pi^*$
orbitals conjugated with the central C=C double bond \cite{Miller}.
TCNE plays a crucial role in molecule-based room-temperature magnets
and spin-injection devices \cite{magnet,TCNESpinInjection}. Given
the local character of the cyano groups and the non-commensurability
of TCNE with the Ag(100) unit cell, a strong stressed adsorption is
to be expected that can give rise to complex patterns
\cite{Weg08,Bed08}. Indeed, we found that TCNE forms a short-range
ordered monolayer exhibiting three non-equivalent adsorption sites.
While STM imaging and STS only show subtle differences in the local
electronic density of states (LDOS) that cannot help to determine
specific adsorption sites, we found that vibrational modes, as
observed in IETS, give quantitative differences in the mode energy
of a particular vibration due to local changes of molecule-substrate
interaction. This inelastic information serves as a reliable input
for comparisons with DFT calculations, permitting structural
determination of the TCNE monolayer.

The experiments were performed in ultrahigh vacuum (UHV) using a
home-built STM operated at $T = 7\, \mbox{K}$. The Ag(100)
single-crystal substrate was cleaned by standard sputter-annealing
procedures, followed by TCNE deposition at room temperature through
a leak valve \cite{Weg08}. After deposition, the sample was
transferred \emph{in situ} to the cryogenic STM. Topography images
were taken in constant-current mode, and STS and IETS were performed
by measuring the differential conductance $dI/dV$ and the 2nd
derivative $d^2I/dV^2$ as a function of the sample bias $V$ by
standard lock-in techniques (modulation $1-10\, \mbox{mV}$ (rms),
frequency $\approx 451\, \mbox{Hz}$) under open-feedback conditions.

%
%

\begin{figure}
\begin{center}
\includegraphics{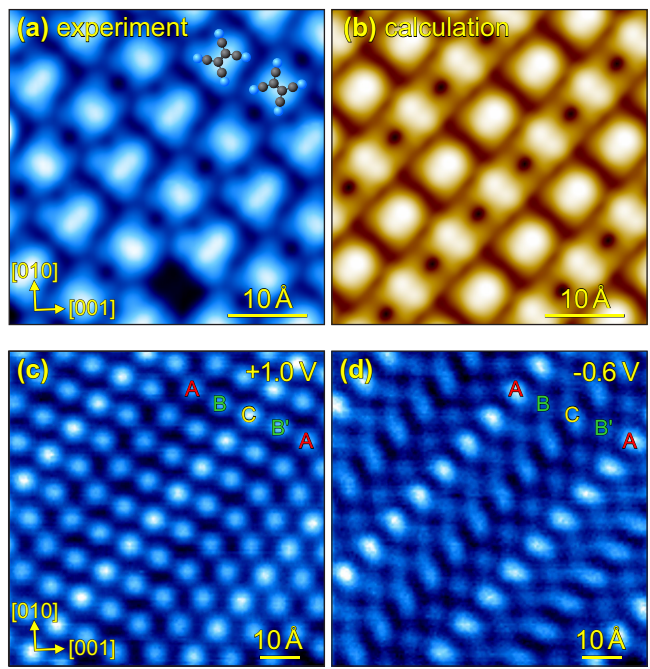}
\caption{\label{fig1} Structural analysis of the TCNE monolayer.
(a)~Highly resolved STM image of the TCNE monolayer on Ag(100). The
intramolecular contrast is dominated by the TCNE LUMO, thus
revealing molecular orientation. (b)~Tersoff-Hamann STM image at
0.1~V derived from the calculated structure (cf.\ Fig.
\ref{Theory10x3} and \ref{Theory}). (c,d)~STM images show the same
area at $+1\, \mbox{V}$ and $-0.6\, \mbox{V}$, respectively. The
strong bias-dependent image contrast indicates distinct variations
of the LDOS for different adsorption sites.}
\end{center}
\end{figure}

Fig.~\ref{fig1} shows STM images of the closed-packed TCNE monolayer
on Ag(100) \bibnote{Local monolayer formation can already be
observed for coverages above 10\% \cite{Weg08}.}. A clear pattern
with local regular order can be observed. Molecules arrange in rows
along the [110] direction with an intermolecular distance of $8.9\pm
0.2\,$\AA, i.e., about three times the Ag(100) surface lattice
constant. Neighboring rows do not form a rectangular arrangement
since the molecules are shifted along the [110] direction in order
to create a dense packing. Fig.~\ref{fig1}a shows a highly resolved
STM image of the monolayer pattern. The molecular shape seen here
mainly reflects the lowest unoccupied molecular orbital (LUMO) of
neutral TCNE and permits determining molecular orientations
\cite{Weg08,Bed08}. Within each row, all TCNEs orient identically
with the C=C double bond being either parallel or perpendicular to
the row. From row to row, the molecular orientation always changes
by $90^\circ$. When the double bond lies along the row direction,
the LUMO shape looks slightly distorted with two of the four cyano
legs appearing less pronounced.

The apparent height of the monolayer exhibits a significant bias
dependence (Fig.~\ref{fig1}c,d). For positive sample bias, each
molecule is seen as an almost round protrusion with only slight
height differences. However, when $V < -0.3\, \mbox{V}$, molecular
shapes are distorted and TCNEs in some rows appear much higher than
in others. Overall, we identify three different heights and mark the
corresponding rows of TCNE with ``A'', ``B'', and ``C'',
respectively. From the highly resolved STM images, we can already
make a first connection between the different electronic properties
and the structure: B molecules always have the C=C double bonds
oriented perpendicular to the row direction, while A and C molecules
always have it aligned parallel.

While this local order is observed throughout the entire monolayer,
we do not observe long-range order. Rather, the pattern is easily
perturbed by local defects or missing molecules. This results in a
variation of combinations of A, B and C rows. In regions with more
defects, the pattern consists of alternating A and B rows only (cf.\
Fig.~\ref{fig1}a), leading to a chevron pattern with a rectangular
$14.4 \times 8.9\, \mbox{\AA}^2 = (5 \times 3)$ unit cell that
contains two TCNEs. In other areas, we observe a larger regular
chevron structure with an A-B-C-B-A... sequence, resulting in a $(10
\times 3)$ unit cell with four molecules. The largest chevron
pattern observed consists of eight molecular rows, as seen in
Fig.~\ref{fig1}c. This pattern reveals a fourth row (B') that,
however, seems to behave similar to B rows. This patterns spans a
unit cell of $(20 \times 3)$ containing eight molecules.

\begin{figure}
\begin{center}
\includegraphics{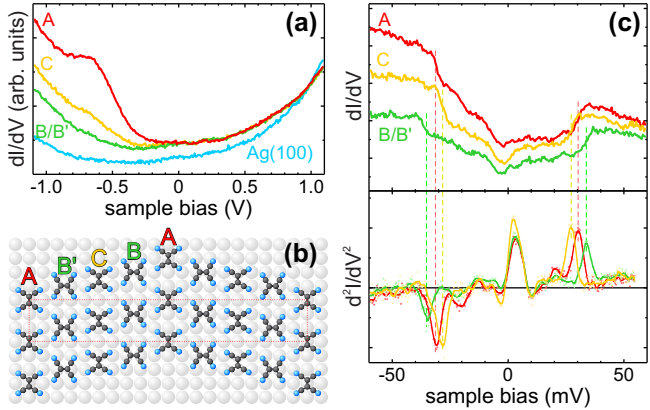}
\caption{\label{fig2} STS and IETS data of TCNE molecules at the
different adsorption sites. (a)~Large-bias spectra reveal LDOS
differences below the Fermi energy. (b)~A rough structural model of
the $(20 \times 3)$ unit cell based only on topography and LDOS
analysis suggests A molecules on top of Ag atoms and C molecules on
bridge sites, which however is only based on comparison with
isolated molecules on Ag(100). Adsorption positions of B and B' are
speculative. (c)~Highly resolved $dI/dV$ and $d^2 I/dV^2$ spectra
reveal a molecular vibration whose exact energy depends on the
adsorption site. For clarity, the Ag-background signal was
subtracted from spectra in (c).}
\end{center}
\end{figure}

In order to better understand the molecular LDOS variations, we have
performed spatially resolved STS measurements on all TCNE molecules
within the monolayer pattern. As expected from the topographical
analysis, we find that all molecules within a row exhibit identical
tunneling spectra. We can therefore summarize all spectroscopic
features by showing a representative spectrum for each row
(Fig.~\ref{fig2}a). Compared to the spectrum on a bare Ag(100)
terrace, all TCNE spectra exhibit only minor differences at positive
bias (i.e., unoccupied states), while we observe a significant
increase in the $dI/dV$ signal at negative sample bias (i.e.,
occupied states). B and B' rows show identical spectra and exhibit a
monotonous increase as we go to larger negative bias. Molecules in C
rows show a similar monotonous LDOS increase, but with a larger
slope. The largest slope is observed for TCNE in A rows, and the
$dI/dV$ spectrum shows an additional broad peak at -0.66~V. This
spectroscopic feature is reminiscent of that observed for isolated
TCNE on Ag(100), where molecules were found to be adsorbed on top of
Ag atoms \cite{Weg08}.

The observed structural and electronic features described up to this
point allow us to propose a rough structural model
(Fig.~\ref{fig2}b) where the rows of TCNE molecules form a locally
commensurate structure on the Ag(100) surface with $(5n \times 3)$
unit cells ($n = 1, 2, 4$). The spectral resemblance with isolated
TCNE leads us to suggest that molecules in row A are likely to be
adsorbed on top of Ag atoms \cite{Weg08}. Further assuming an
arrangement of molecules A-B-C-B'-A along a straight line, we find
that for C molecules the bridge site is a probable adsorption
position with high symmetry. However, B and B' molecules would then
lie in a position of very low symmetry. Within our experimental
accuracy, we cannot rule out that these molecules relax laterally to
a bridge or hollow site, both being only about 0.7~{\AA} away.
Consequently, our experimental structure model requires comparison
with calculations which are rendered difficult due to the lack of
quantifiable spectroscopic features.

The situation changes altogether when we take a look at the
vibrational structure of the TCNE molecules by performing STS with
high energy resolution (Fig.~\ref{fig2}c). Again, we find that all
molecules within a specific row exhibit identical spectra. In all
cases $dI/dV$ spectra show pronounced step-like features at about
$+30$ and $-30\,$meV with a conductance change of 2-3\%. This is a
well-known signature of IETS (Fig.~\ref{fig2}c, top) \cite{Sti98}.
The energy of this feature is in good agreement with reported values
of the in-plane rocking mode as well as the out-of-plane wagging
mode of TCNE \cite{Mic82}, and it has been observed via STM-IETS for
TCNE in various local environments \cite{Weg09,Choi10}. Upon closer
inspection, we find that the exact IETS energy depends on the
location of TCNE within the monolayer pattern. This can be seen
clearly in the $d^2 I/dV^2$ spectra (Fig.~\ref{fig2}c, bottom).
Molecules exhibit a mode energy of $30.7 \pm 0.5\,$meV in row A,
$34.3 \pm 1.0\,$meV in row B, and $27.7 \pm 0.4\,$meV in row C. The
error bars are much smaller than the frequency shifts and the actual
measured frequencies for single measurements never overlapped among
different types of molecules. These data are based on a statistical
analysis of 32 molecules. Hence, the observed three different mode
energies are clearly resolvable and distinguishable. Stiffening or
softening of vibrational modes can be caused by local variations of
intermolecular or molecule-substrate interactions. Therefore, the
observed energies serve as quantitative fingerprints that can help
to identify the three non-equivalent adsorption sites of TCNE within
the monolayer. Despite both sitting on bridge sites, the B and C
conformations are different because the molecular C--C axis in C
molecules aligns along the row, while in B rows it is perpendicular
to the row. Hence, an individual B molecule is $90^\circ$ rotated
with respect to an individual C molecule. As a consequence, the
reconstructed Ag layer below B and C molecules is different,
affecting both local electronic structure and mode frequency.


\begin{figure}
\begin{center}
\includegraphics{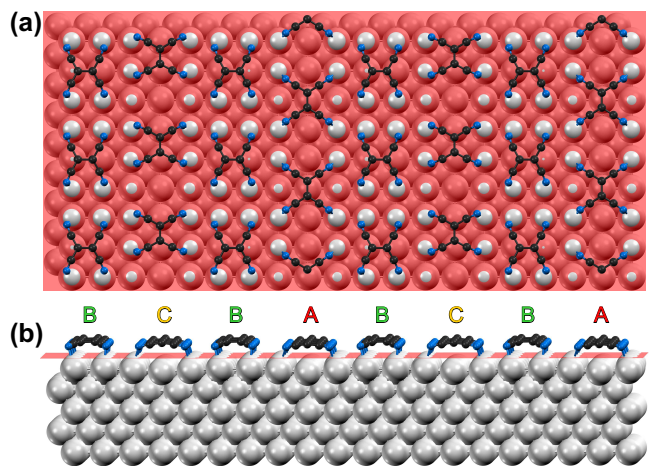}
\caption{\label{Theory10x3}DFT structure of the $(10 \times 3)$ TCNE
monolayer on Ag(100), in top view (a) and side view (b). A
transparent plane was added to emphasize the vertical relaxation of
Ag atoms in the first layer.}
\end{center}
\end{figure}

This adsorption-site determination can be achieved by comparing our
experimental findings with DFT calculations. We show that comparing
only the electronic structure is inconclusive, whereas the added
information from IETS is decisive to permit the correlation between
vibrational modes and the underlying geometrical structure. We first
perform a structural analysis by calculating a ($10\times3$) unit
cell containing four molecules. The substrate was modeled using five
Ag substrate layers. The first two layers and the molecules were
relaxed until forces on the atoms were less than 0.01 eV/\AA. We
used the {\tt vasp} code and the PBE approximation for the exchange
and correlation functional \cite{vasp,vasp2,PBE}.
Fig.~\ref{Theory10x3} shows the geometry of the simulated TCNE
monolayer on Ag(100). A transparent plane was added to emphasize the
degree of distortion of Ag atoms in the surface layer due to
molecule-substrate interactions. Type A molecules sit on top of a
silver atom with elastic substrate-mediated intermolecular
interactions among A molecules since the substrate distortion aligns
parallel to the molecular rows of the same species. Type B molecules
are on bridge sites and are completely surrounded by high-lying Ag
atoms. Type C molecules also adsorb on bridge sites but with an
orientation similar to A-type molecules, which also leads to an
aligned substrate distortion parallel to the molecular row. Thus,
the local adsorption environment is indeed different for all three
types of TCNE molecules. All molecules are adsorbed via the N--Ag
local interactions, and the arrangement provides denser packing. The
simulated Tersoff-Hamann image (Fig.~\ref{fig1}b) is in very good
agreement with the experimental STM topography at small sample bias
\cite{Tersoff-Hamman}. Calculations of the density of states
projected onto the molecular orbitals (not shown), demonstrate that
the LUMO of all molecules is broadened and shifted slightly below
the Fermi energy. While this explains the LUMO character in STM
images at small sample bias, the differences in the electronic
structure between A, B and C sites are too small to account for the
experimentally observed image contrast at larger bias (Fig.~1c,d) as
well as the STS differences seen in Fig.~\ref{fig2}a. Thus, a
comparison between the experimentally derived and the calculated
structural model via electronic structure is not conclusive, likely
due to the reliance of the simulations on the non-physical Kohn-Sham
orbitals.

\begin{figure}
\begin{center}
\includegraphics[width=0.4\textwidth]{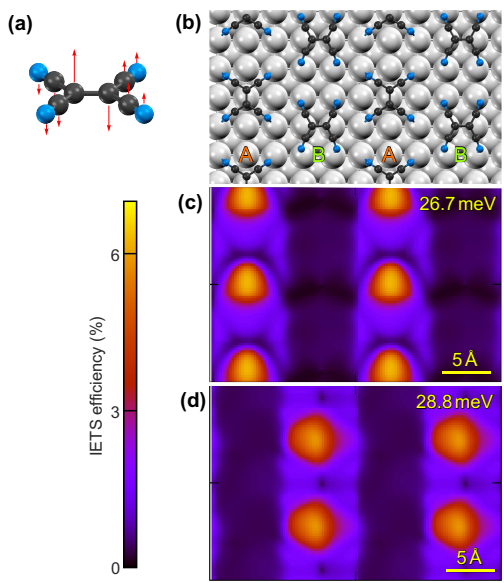}
\caption{\label{Theory} DFT calculations and IETS simulations of the
TCNE monolayer on Ag(100). (a)~Scheme of the out-of-plane wagging
mode that is excited in STM-IETS. (b)~Atomic scheme of the ($5
\times 3$) structure showing types A and B molecules in different
orientations on top and bridge sites, respectively. (c,d)~IETS
efficiency in \% of tunneling electrons plotted for the out-of-plane
wagging mode of type A (26.7~meV) and B (28.8~meV) molecules.}
\end{center}
\end{figure}

On the other hand, the vibrational structure simulations do not
suffer from these deficiencies. For the analysis of vibrational
modes, we restrict our discussion to the smallest observed
($5\times3$) unit cell containing only molecules A and B
(Fig.~\ref{Theory}a). The IETS simulation is performed using a
many-body perturbation extension of the Bardeen tunneling theory
\cite{Lorente2000,Lorente2004}. Unfortunately, DFT calculations of
vibrational energies of adsorbed molecules on a surface are
generally shifted and at best within 10\% of the experimental
values. Hence, while absolute frequency values are not reliable, the
frequency difference between modes is very accurate
\cite{Bocquet2006}. Our IETS simulations predict that only one mode
dominates the signal within the experimental energy range. This is
the out-of-plane wagging mode illustrated in Fig.~\ref{Theory}c. Two
different frequencies are found for molecules A and B. The
out-of-plane wagging mode of type A molecules is located at $E_A =
26.7\,$meV, while that of type B molecules is found at $E_B =
28.8\,$meV. The maximum fraction of computed inelastic electrons is
7\%, whereas all other modes yield inelastic fractions below 1\%.
The direction and magnitude of the energy shift $E_B - E_A =
2.1\,$meV are in good agreement with the experimentally determined
shift of $3.7\pm 1.2\,$meV. The frequency shift of the other
molecular modes does not always follow this trend. Especially, the
second possible candidate, the in-plane rocking mode, is found at
30.8~meV for molecule A and at 28.3~meV for molecule B in the
simulations, i.e., the shift is opposite to the experimental
observation. Using DFT-D2 to account for dispersion forces does not
alter this conclusion \bibnote{Neglecting dispersion corrections can
lead to wrong vertical molecule-substrate distances and
intramolecular distortions
\cite{Romaner2009,Bluegel2009,Scheffler2012}. Using the DFT-D2
approach as coded in VASP \cite{Grimme2006}, the N atoms move
0.03~{\AA} and the C=C double bond moves 0.25~{\AA} closer to the
surface. However, the vibrational modes exhibit only minor
softening. In particular, the wagging mode is shifted by 2~meV, well
within the error bar of the calculations.}. The simulations hence
show that only the out-of-plane wagging mode is excited in the
experiment.

We have performed an equivalent calculation for a fictitious CBC
structure in order to prove that indeed our geometrical assignment
is correct. The CBC structure is energetically less favorable than
the ABA structure and indeed, it is not found experimentally
\bibnote{Our calculations show that a molecule adsorbed with its
$C=C$ axis on a top site (such as the A molecules) are 0.6 eV more
bound to the substrate than molecules on a bridge site (such as B
and C molecules). Hence stable structures contain A molecules in
their periodic pattern.}. Nevertheless, due to the locality of
vibrational properties, we expect that the CBC structure captures
the frequency shifts of C with respect to B molecules, although we
emphasize that a quantitative comparison with the ABA calculations
is not valid. Despite this, B molecules show a wagging mode
frequency of 29.2 meV in very good agreement with the value of the
ABA structure. Furthermore, we find that the wagging mode of C
molecules is 1.9 meV lower in energy than that of B molecules, i.e.,
the trend is in agreement with the experimental finding. Thus, the
mode frequency is not only determined by the adsorption site but
also by the local molecular environment.

Due to the localization of the vibrations to each molecule, the IETS
signal is also well localized. This permits detecting the different
molecular adsorption sites. Localization is a general feature of
intramolecular modes. In fact, it leads to very weakly dispersing
optical-like phonons in molecular adlayers that can be individually
excited within a molecule \cite{Bocquet2009}. Fig.~\ref{Theory}c
shows a map of conductance change with bias when the mode at $E_A$
is excited. The IETS simulations clearly reflect the spatial
distribution of type A molecules with a maximum IETS signal at the
molecular center, in agreement with the experimental findings. When
the mode at $E_B$ is excited, it is localized to type B molecules,
as shown in Fig.~\ref{Theory}d. Thus, we find very good agreement
between the experimental IETS results and the calculated vibrational
structure and IETS simulations, which permits to confirm the
DFT-calculated structure conclusively. Particularly, we conclude
that type A molecules adsorb on top and type B and C molecules on
bridge sites of the Ag(100) surface.

This proof-of-principle study shows that IETS combined with DFT can
be used to discriminate between non-equivalent molecular adsorption
environments in a dense complex molecular monolayer. The high
chemical sensitivity of IETS enables the detection of small
variations in molecular environments that easily lead to meV
spectroscopic changes, i.e., well within the typical IETS energy
resolution, while the high spatial resolution of IETS displays
intramolecular localization when intrinsic molecular modes are
excited. These unique properties combined with DFT reveal hidden
geometrical structure not attainable by the usual structural
methods.

%
%
\begin{acknowledgement}

This work was supported by the Director, Office of Science, Office
of Basic Energy Sciences of the US Department of Energy under
contract no. DE-AC02-05CH11231 (STM instrumentation development and
measurements), and by the Deutsche Forschungsgemeinschaft (DFG)
project WE 4104/2-1 (numerical simulations and analysis). D.W.
acknowledges support by the Alexander von Humboldt Foundation (data
acquisition) and the North-Rhine Westphalian Academy of Sciences and
Arts (data analysis). N.L. is supported by the ICT-FET Integrated
Project AtMol (http://www.atmol.eu) (IETS simulation code
development).

\end{acknowledgement}

\bibliography{TCNE-Monolayer}

\begin{mcitethebibliography}{56}
\bibitem[Not()]{Note-1}
Local monolayer formation can already be observed for coverages above 10\%
  \cite{Weg08}.\relax
\mciteBstWouldAddEndPunctfalse
\mciteSetBstMidEndSepPunct{\mcitedefaultmidpunct}
{}{\mcitedefaultseppunct}\relax
\EndOfBibitem
\bibitem[Not()]{Note-3}
Our calculations show that a molecule adsorbed with its $C=C$ axis on a top
  site (such as the A molecules) are 0.6 eV more bound to the substrate than
  molecules on a bridge site (such as B and C molecules). Hence stable
  structures contain A molecules in their periodic pattern.\relax
\mciteBstWouldAddEndPunctfalse
\mciteSetBstMidEndSepPunct{\mcitedefaultmidpunct}
{}{\mcitedefaultseppunct}\relax
\EndOfBibitem
\end{mcitethebibliography}

\begin{tocentry}
  \includegraphics{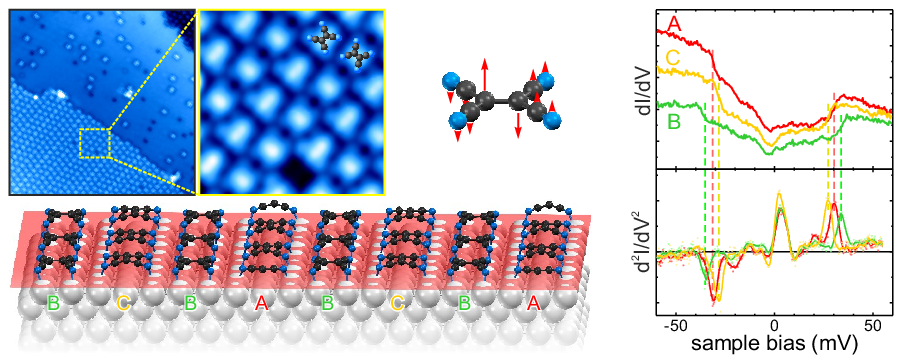}
\end{tocentry}

\end{document}